# Correlation Between DNA Double-Strand Break Distribution in 3D Genome and Ionizing Radiation-Induced Cell Death

Ankang Hu[1,2], Wanyi Zhou[1,2], Xiyu Luo[3], Rui Qiu[1,2,*], and Junli Li[1,2,*]

[1] Department of Engineering Physics, Tsinghua University, Beijing, China

[2] Key Laboratory of Particle & Radiation Imaging, Tsinghua University, Ministry of Education, Beijing, China

[3] Institute of Fluid Physics, China Academy of Engineering Physics, Mianyang, China

*Authors to whom any correspondence should be addressed

Email: lijunli@mail.tsinghua.edu.cn and qiurui@tsinghua.edu.cn

**Abstract.** The target theory is the most classical hypothesis explaining radiation-induced cell death, yet the physical or biological nature of the "target" remains ambiguous. This study hypothesizes that the distribution of DNA double-strand breaks (DSBs) within the 3D genome is a pivotal factor affecting the probability of radiation-induced cell death. We propose that clustered DSBs in DNA segments with high interaction frequencies are more susceptible to leading to cell death than isolated DSBs. Topologically associating domains (TAD) can be regarded as the reference unit for evaluating the impact of DSB clustering in the 3D genome. To quantify this correlation between the DSB distribution in 3D genome and radiation-induced effect, we developed a simplified model taking into account the DSB distribution across TADs. Utilizing track-structure Monte Carlo codes to simulate the electron and carbon ion irradiation, we calculated the incidence of each DSB case across a variety of radiation doses and linear energy transfers (LETs). Our simulation results indicate that DSBs in TADs with frequent interactions (case 3) are significantly more likely to induce cell death than clustered DSBs within a single TAD (case 2). Moreover, case 2 is significantly more likely to induce cell death than isolated DSBs (case 1). The curves of the incidence of case 2 and case 3 versus LETs have a similar shape to the radiation quality factor (Q) used in radiation protection. This indicates that these two cases are also associated with the stochastic effects induced by high LET irradiation. Our study underscores the crucial significance of the 3D genome structure in the fundamental mechanisms of radiobiological effects. The hypothesis in our research offers novel perspectives on the mechanisms that regulate radiobiological effects. Moreover, it serves as a valuable reference for the establishment of mechanistic models that can predict cell survival under different doses and LETs.

## Introduction

Radiotherapy is one of the main cancer treatment methods, where the induction of cell death via radiation plays a crucial role. The underlying mechanism by which radiation induces cell death forms the cornerstone of both the theory and practice of radiotherapy. The survival fraction of mammalian cells following irradiation at a specific dose is effectively captured by the survival fraction curve(*1, 2*). Characteristically, for low linear energy transfer(LET) irradiation, this curve exhibits a "shoulder" at lower dose ranges, transitioning to a near-linear relationship at higher dose ranges(*3, 4*), but for high-LET irradiation, the shoulder typically tends to diminish. The survival fraction curve is the fundamental data for the clinical application of radiotherapy(*5, 6*).

Several theories have been proposed to elucidate the survival fraction curve, and corresponding models have been developed based on these theories to predict cell death resulting from irradiation at specific doses and LETs. The "target theory" ranks among the most influential of these (*7*). It postulates that cells have one or several crucial "targets". When these targets are damaged by irradiation, which is referred to as a "hit" or "lesion", it results in cell death. The "target theory" serves as the foundation for the vast majority of radiobiological theories.

The Linear-Quadratic (LQ) model(*8*) is the most widely utilized model in clinical practice, effectively quantifying the survival fraction within a range of several Gy. This model, as represented by Eq 1, suggests that the destruction of a target requires at least two lesions, which may be inflicted by either a single particle or multiple particles. The contribution of these lesions is quantified by the linear and quadratic terms within the model.

$$-\ln S = \alpha D + \beta D^2 \tag{1}$$

Where $S$ represents the survival fraction, $D$ denotes the radiation dose, and $\alpha$ (unit: Gy$^{-1}$) and $\beta$ (unit: Gy$^{-2}$) are two parameters that are specific to cell types. The ratio $\alpha/\beta$, a critical parameter in radiotherapy, exhibits variability among different cell types, with typical values ranging from 2 Gy to 10 Gy.

"Structure determines function" is a fundamental principle in biology. The exact physical and biological entities of the "target" and "hit" remain an area of ongoing investigation.

Initially, Chadwick and Leenhouts postulated that the double-stranded DNA has been identified as the primary "target" with the strand breaks caused by radiation being considered as "damage"(*9*). The double-strand break (DSB) induced by a single radiation particle constitutes the first-order term in Eq 1. The breakage of two opposite strands by separate particles, which subsequently combine to form a double-strand break, contributes to the second-order term. However, ample evidence indicates that the number of DSBs is directly proportional to the radiation dose(*10–14*), rather than adhering to the linear-quadratic formula as outlined in Eq 1. The occurrence of DSBs alone is insufficient to account for the formation of the cell survival curve.

In recent years, the chromosome structure with a larger size has been regarded as the key "target" in radiobiology. Researchers have differentiated DSBs into two primary categories: isolated DSBs and clustered DSBs(*15–17*). Isolated DSBs represent singular events, while clustered DSBs (not referred to the lesion with more than two single strand breaks within 10 bp) are characterized by the occurrence of more than two DSBs within a defined "domain". The "domain" is referred to as a "giant loop" in chromatin, which spans a region of 1-3 megabase pairs (Mbp) of DNA. Through the fitting of parameters against cell survival curve data, investigators have discovered that the probability of cell death resulting from complex DSBs is several orders of magnitude greater than that caused by isolated DSBs. Additionally, radiobiological effect models that account for high LET irradiations, such as the microdosimetric kinetic (MK) model(*18, 19*) and the local effect model (LEM)(*20*), both propose the presence of a "domain" within the nucleus, approximately hundreds of nanometers in radius and encompassing about 2 Mbp of DNA (*21*). However, recent advancements in 3D genome mapping have revealed the intricate structure of chromosome folding within the nucleus, casting doubt on the concept of the "giant loop" as a distinct physical or biological entity within the cell(*22, 23*). Further research is imperative to investigate the specific domains that regulate radiobiological effects, aiming to elucidate the complex relationship between radiation-induced damage and cellular response.

In this study, we hypothesize that the probability of DSBs leading to cell death is related to their distribution within the 3D genome. To quantify this hypothesis, we have developed a simplified model that takes into account the distribution of DSBs across topologically associating domains(TADs). Subsequently, we investigate the correlation between the distribution of DSBs across TAD and cell death induced by irradiation at various doses and LETs using track-structure Monte Carlo simulations. The simulation focuses on electrons and carbon ions to represent low and high LET irradiations. Our hypothesis provides insights into the mechanisms of radiobiological effects and offers a valuable reference for developing mechanistic models that predict cell survival across a range of doses and LETs.

## Materials and Methods

### Hypothesis

In eukaryotic cells, chromatin is not randomly dispersed within the nucleus, rather, it is organized into a sophisticated hierarchical architecture, commonly referred to as the 3D genome(24, 25). Even genomic segments that are remote from each other in the linear sequence can engage in interactions through this 3D genomic structure, facilitating complex regulatory mechanisms. Numerous chromatin conformation capture assays, including Hi-C, have been utilized to explore the architecture of the 3D genome, revealing structural elements such as chromatin domains, compartments, and TADs(26). Genes that exhibit frequent interactions are often in close spatial proximity and are associated with the same proteins, which can be detected through cross-linking in Hi-C experiments. The 3D genome structure governs the spatial proximity and interactions between DNA segments, and it is likely to be closely associated with the biological effects induced by radiation. Furthermore, TADs serve as functional units in the DNA damage response (27). For instance, they play a pivotal role in the loop extrusion process during double-strand break repair (28). Additionally, there may be a correlation between TADs and the effects induced by radiation.

We propose a hypothesis that DSBs that are clustered in the 3D genome, where the DNA segments containing DSBs exhibit high frequencies of interaction, the probability of incorrect repair leading to structural variations is significantly elevated compared to DSBs within segments that engage in infrequent interactions. As a result, the likelihood of cell death resulting from the clustered DSB scenario is considerably higher than that observed in the less frequently interacting segment scenario. Additionally, within the context of the clustered DSB scenario, the proximity of the two DNA segments in the genome is a critical factor: if the segments are in close genomic proximity, the occurrence of small-segment structural variation is more probable. Conversely, a large genomic distance between the segments is more likely to result in large-segment structural variations. We also suggest that the various types of structural variations may have differing probabilities of inducing cell death. Furthermore, TADs can be conceptualized as the reference unit for evaluating the impact of DSB clustering in the 3D genome on radiation-induced cellular responses. The diagram of our hypothesis is shown in Figure 1.

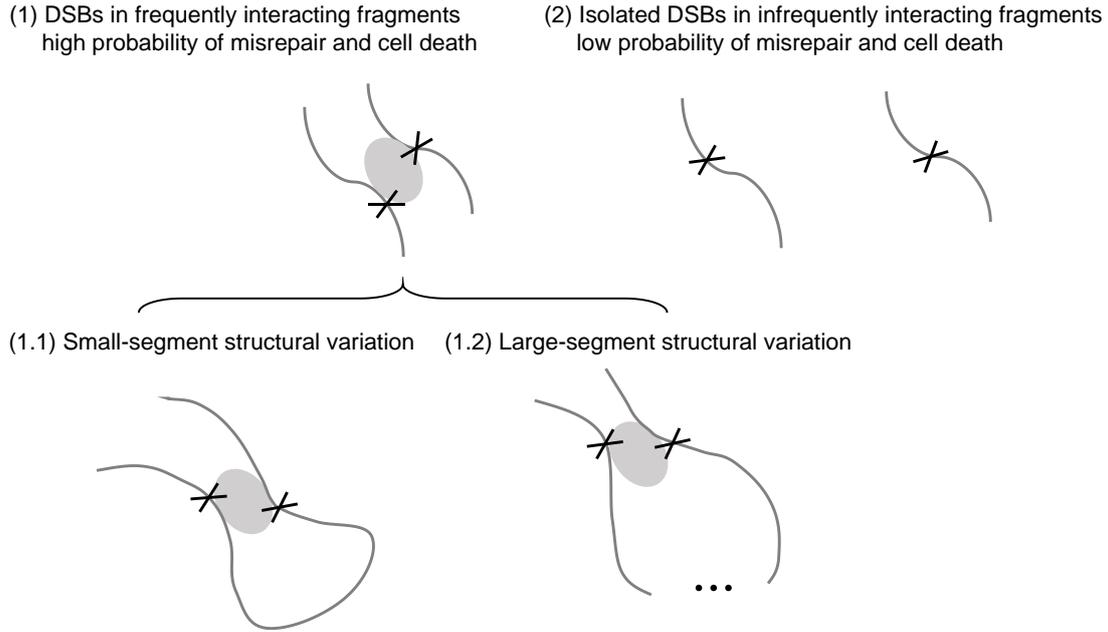

Figure 1. Diagram of our hypothesis

**Quantifying the hypothesis through a mathematical model**

To quantitatively assess our hypothesis, we have developed a simplified mathematical model. Establishing a quantitative relationship between interaction frequency and the probability of cell death is quite challenging. To address this challenge, we have categorized the distribution of DSBs in terms of TADs, since TADs are regarded as the fundamental units in the 3D genome structure. (*23*, *29*, *30*). We have categorized the distribution of DSBs in the 3D genome into three distinct cases according to the interaction matrix obtained from Hi-C data, as depicted in Figure 2. For the sake of simplicity, our model assumes the same cell death probability within the same case, $p_1$ for case 1, $p_2$ for case 2 and $p_3$ for case 3. In our classification system:

- Case 1 represents an isolated DSB, where a single TAD contains one DSB, and TADs that typically interact with this TAD are free of DSBs. This corresponds to scenario (2) in Figure 1.

- Case 2 is defined by clustered DSBs within a single TAD, where a single TAD harbors two or more DSBs. This corresponds to scenario (1.1) in Figure 1.

- Case 3 involves DSBs in TADs with frequent interactions (each TAD contains one or more DSBs, and these TADs interact frequently with each other). This corresponds to scenario (1.2) in Figure 1.

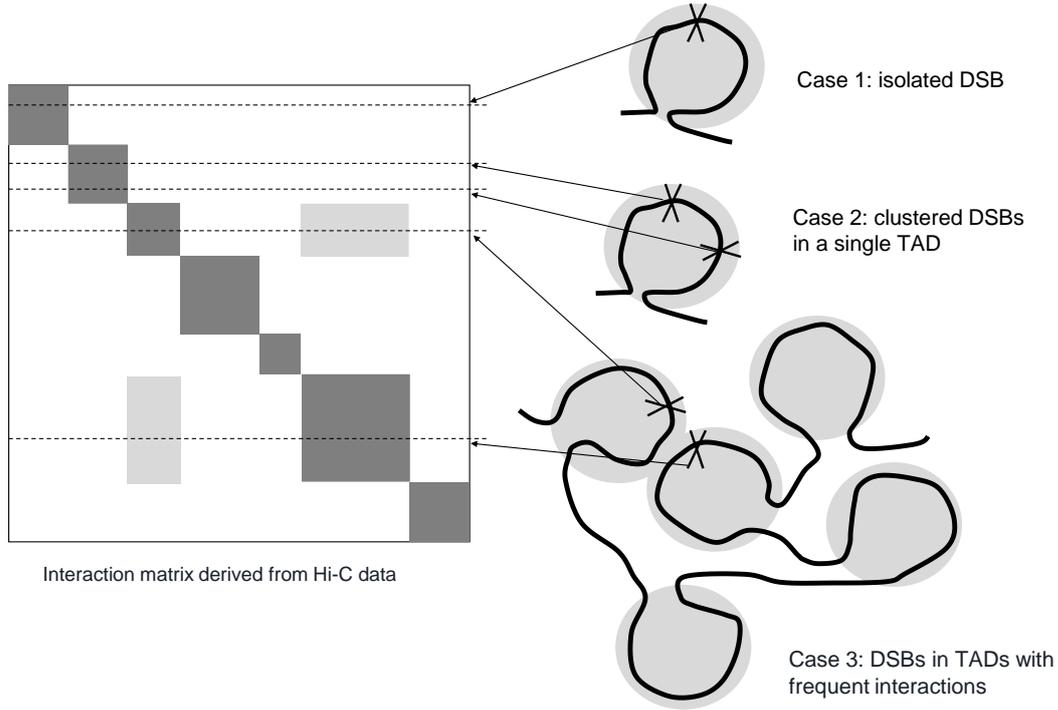

Figure 2. Schematic diagrams of the three cases in terms of geometry and Hi-C data. The dotted lines in the interaction matrix correspond to the positions of DSBs in the genome. The light grey colored blocks represent the frequent interactions between two TADs.

The incidence of each case within a cell irradiated with a specific dose, $D$, can be denoted as $n_1(D)$ for Case 1, $n_2(D)$ for Case 2, and $n_3(D)$ for Case 3. Each case has a probability of $p_1$, $p_2$, and $p_3$, respectively, of resulting in a lethal event. The total number of lethal events can be determined using the formula provided in Eq 2.

$$n(D) = p_1 n_1(D) + p_2 n_2(D) + p_3 n_3(D) \qquad (2)$$

Adopting the same assumptions employed in deriving the survival fraction model from the "target theory", the number of lethal events is approximately Poisson-distributed. The survival fraction can then be computed using the formula provided in Eq 3.

$$S = \exp\left[-n(D)\right] \qquad (3)$$

Consequently, we can forecast the survival fraction by the number of the three cases, utilizing the predictive formula detailed in Eq 4.

$$-\ln S = p_1 n_1(D) + p_2 n_2(D) + p_3 n_3(D) \qquad (4)$$

The functions $n_1(D)$, $n_2(D)$ and $n_3(D)$ are determined by track-structure Monte Carlo simulations, which are elaborated in the subsequent section. The probabilities $p_1$, $p_2$, and $p_3$ are determined by specific cell types.

**Track-structure Monte Carlo simulation study**

We utilized Geant4-DNA, a track-structure Monte Carlo simulation code previously developed, to compute the energy deposition events within the nucleus(*31*, *32*). Subsequently, the data pertaining to these energy deposition events were imported into Python script for the analysis of DNA damage.

To investigate the distribution of DNA damage within the 3D genome, we utilized a nuclear model of the IMR90 cell. This model delineates TADs without lamina-associated domains (LADs) and was developed by Ingram et al. based on Hi-C data using the code G-NOME (*33*). The radius of the nucleus is 5.0 μm. The occupancy volume is 0.15. This nuclear model represents TADs as spherical units, with the positions and radii of each sphere determined by a 3D genome modeling algorithm (*22*, *23*). There are 15,282 TADs, which contain 6.07 Gbp of DNA. The mean radius of the TAD spheres is 89.12 nm, with a standard deviation of 41.29 nm. The mean DNA content of a TAD is 397 kbp, with a standard deviation of 840 kbp. The distributions of the TAD radii and DNA content are presented in the supplementary materials (Figure S1 and S2). Additionally, the model provides information on the TADs that are known to interact frequently with a given TAD. The criteria are for identifying frequently interacting TAD developed by Chrom3D based on a statistical testing method (*22*). Significant interactions are identified through the utilization of the NCHG module within Chrom3D. This module calculates a *P* value based on the probability of observing a given number of contacts conditional on the total number of contacts for both regions involved, as well as the total number of contacts, using a non-central hypergeometric distribution. The interactions are then selected with a false-discovery rate set at 1% (*34*). We use a single nucleus in the simulation because the sizes of TADs and the relative contacts among TADs are kept for all G-NOME-generated nuclear models with the same Hi-C data. In each simulation, we randomly rotate the coordination of energy deposition events by a specific angle to mimic the various angles at which the cell is exposed to radiation.

We used Geant4 version 11.2 for the physics simulations. The physics list is the same as that of G4EmDNAPhysics_option2, but we changed the energy limit of the G4DNARuddIonisationExtendedModel up to 500 MeV/u to achieve low LET carbon ion simulations. This change may result in less precise simulations because the default limit is 300 MeV/u. We define the simulation world as a water cube with an edge length of 120 μm. The nuclear model is positioned at the center of the cube. The ion source is represented as a plane situated 5.5 μm away from the center of the cube to ensure all TAD spheres are in front of the source. The starting points of the particles are uniformly and randomly sampled from this plane, and their directions are set to be parallel to the normal vector of the plane.

For the simulation of each particle energy, we establish different physic simulation groups to approximate a dose range from 1.0 to 10.0 Gy. For each of these physics simulation groups, the physics simulations are executed 10 times, with a fixed number of simulated particles in each execution. The number of particles is determined according to the dose delivered to a sphere of 5 μm radius positioned at the center of the space. Specifically, we choose the particle number corresponding to a dose close to 1.0 Gy. For each simulation run, the analysis of DSBs is performed 100 times. In each DSB analysis instance, the coordinates of the energy deposition events are rotated by a random angle. The results of the DSB analysis reflect the estimates of average incidences of the cases induced by the specific dose (defined by the centered sphere). We assume that these incidences adhere to a Poisson distribution. In a Poisson distribution, the variance is equal to the mathematical expectation. Our results inherently incorporate information about the associated variability.

In the analysis of DNA damage, energy deposition events that occur within TADs are recorded. To translate these energy deposition events into strand breaks, two conditions are implemented. First, for the energy deposition events occurring within the TAD spheres, we randomly select 14.1% of the events that hit the DNA (*33*). This percentage is defined as $p_{DNA}$. We have set three different values of $p_{DNA}$ (0.141, 0.17 and 0.21) to show the impact of these parameters on the results.

Second, an energy range probability function is applied, which varies from 0 at 5 eV to 1 at 37.5 eV(*12, 35, 36*), indicating the likelihood of an energy deposition event resulting in a strand break. An energy deposition event is considered a strand break if it meets both conditions and is then randomly assigned to either strand 1 or strand 2 of the DNA double helix with equal probability. DSBs are defined as two or more strand breaks that occur on opposite strands and are within a distance of 3.2 nm or less from each other, which is approximately equivalent to 10 base pairs. All parameters for this setup are consistent with those used by Ingram et al. (*33*). The Density-Based Spatial Clustering of Applications with Noise (DBSCAN) algorithm is employed to analyze the distribution and clustering of DSBs. Then the incidences of three cases are counted for subsequent analysis. The Python scripts used for the analysis of DNA damage are available at the URL (https://github.com/hak1996/DSB_3D_genome).

We use approximate formulas for describing the incidence of the three cases vs dose as shown in Eq 5. For case 1 ($n_1$), we employ a second-order term to adjust the value at high dose levels. For case 2, we take into account the events induced by single particles ($n_{2,s}$) and multiple particles ($n_{2,m}$). For case 3 ($n_3$), we utilize a fourth-order polynomial.

$$
\begin{aligned}
n_{DSB}(D) &= k_{DSB}D \\
n_1(D) &= a_1 D - b_1 D^2 \\
n_{2,s}(D) &= a_{2,s} D - b_{2,s} D^2 \\
n_{2,m}(D) &= b_{2,m} D^2 - c_{2,m} D^3 \\
n_2(D) &= n_{2,s}(D) + n_{2,m}(D) \\
n_3(D) &= a_3 D + b_3 D^2 - c_3 D^3 + d_3 D^4
\end{aligned}
\tag{5}
$$

Where $k_{DSB}$, $a_1$, $b_1$, $a_{2,s}$, $b_{2,s}$, $b_{2,m}$ $c_{2,m}$, $a_3$, $b_3$, $c_3$ and $d_3$ are parameters obtained by fitting the results of Monte Carlo simulations.

We simulate 0.9 MeV electrons to represent low LET irradiation and C-12 ions with energies ranging from 2.98 MeV/u to 354.1 MeV/u to represent high LET irradiation. The relationship between the energy of the ion and the LET is calculated based on the data provided by Furusawa et al. (*37*).

**Acquisition of Cell Type-Specific Parameters**

The simulation outcomes from electron irradiation are analyzed to derive the formulas for three distinct cases ($n_1(D)$, $n_2(D)$, and $n_3(D)$). We employ polynomial functions to approximate these formulas within the dose range. Subsequently, the probabilities of cell death for the three cases ($p_1$, $p_2$ and $p_3$) are determined by fitting a photon-irradiation curve and an additional high LET irradiation curve (ranging from approximately 70 - 90 keV/μm, according to the papers of the data source). We employ the nonlinear least-squares fitting methods provided by MATLAB to estimate these parameters. During the fitting process, the lower bound of these three parameters is set to zero. Utilizing the obtained probabilities, we then calculate the survival fraction of cells irradiated by C-12 ion with energies ranging from 2.98 MeV/u to 354.1 MeV/u.

**Results and Discussions**

**Dose-dependency of the incidence of three cases**

The simulation results for electron and 75 keV/μm C-12 ion irradiation are presented in Figure 3. The parameters of Eq 5 and the estimated 95% confidence interval (95% CI) are listed in Table 1.

For electron irradiation, almost all instances of case 3 are induced by multiple particles. The value of $a_3$ is zero, so it is not shown in Table 1. For C-12 ion irradiation, the value of $d_3$ is so small that it can be neglected. It is not presented in the table.

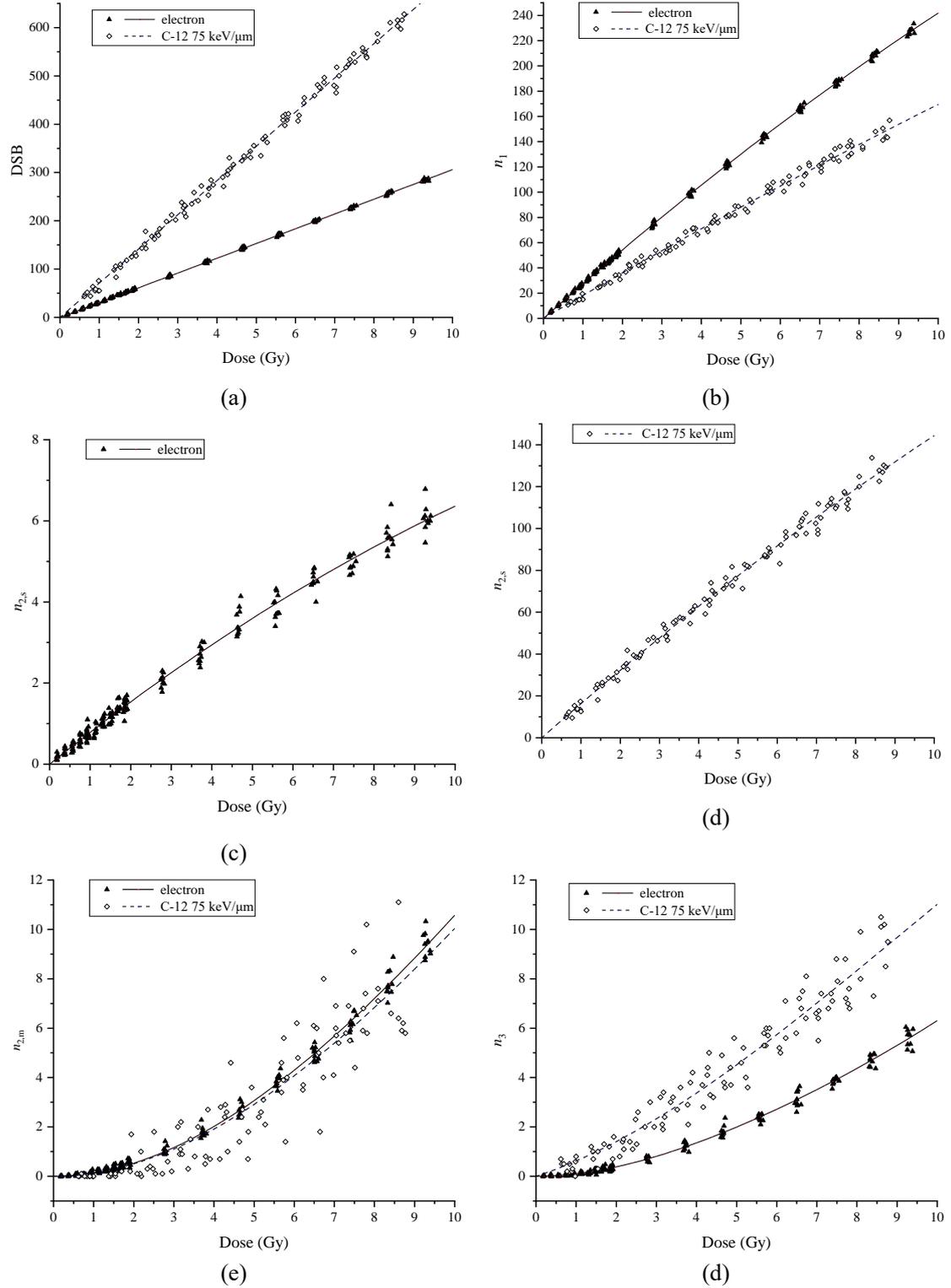

Figure 3. Incidence of events induced by electron and 75 keV/μm C-12 ion. (a) DSB; (b) isolated DSB (case 1); (c, d) clustered DSB in a single TAD (case 2) induced by one particle; (e) clustered DSB in a single TAD (case 2) induced by multi particles; (f) DSBs in TADs with frequent interactions (case 3).

Table 1. Parameters of formulas describing the incidence of three cases induced by electron and 75 keV/μm C-12 ion (with 95% CI)

| Parameters | electron | C-12 75 keV/μm |
|---|---|---|
| $k_{DSB}$ | 30.59 (30.54 - 30.64) | 70.85 (70.42 - 71.29) |
| $a_1$ | 27.8 (27.70 - 28.00) | 18.28 (17.75 - 18.80) |
| $b_1$ | 0.3665 (0.3462 - 0.3868) | 0.1327 (0.005536 - 0.2100) |
| $a_{2,s}$ | 0.8009 (0.7785 - 0.8233) | 16.54 (16.03 - 17.04) |
| $b_{2,s}$ | 0.01646 (0.01343 - 0.01949) | 0.2097 (0.135 - 0.2844) |
| $b_{2,m}$ | 0.1396 (0.1338 - 0.1454) | 0.1326 (0.09383 - 0.1713) |
| $c_{2,m}$ | 0.003387 (0.002682 - 0.004092) | 0.003222 (-0.001995 - 0.008438) |
| $a_3$ | / | 0.5359 (0.2819 - 0.7899) |
| $b_3$ | 0.1081 (0.0954 - 0.1208) | 0.09028 (0.000624 - 0.1799) |
| $c_3$ | 0.006831 (0.003222 - 0.01044) | 0.003374 (-0.004178 - 0.01093) |
| $d_3$ | 0.0002325 (-0.0000163 - 0.0004813) | / |

The experimental results indicate that the DSB yield for low LET irradiation falls within the range of 30 to 50 Gy$^{-1}$ (*12*, *14*, *33*). Our results are in agreement with the reported DNA DSB yields for low LET radiation.

For electron irradiation, case 2 induced by multiple particles and case 3 might contribute to the "shoulder" observed in the cell survival fraction curve at low doses. As the dose increases, the incidences of case 2 and case 3 tend to follow straight lines, corresponding to the "straight line" portion of the survival fraction curve observed at high doses. For high LET irradiation, the instances of case 2 induced by multiple particles are much fewer than those induced by a single particle. Additionally, the curve of case 3 more closely approximates a straight line. These results correspond to the survival fraction curve of high LET irradiation.

The simulation parameter $p_{DNA}$ significantly influences the yield of DSBs and the incidence of the three cases. The results of different $p_{DNA}$ values are shown in the supplementary materials (Figure S3 and Table S1). The parameter is determined by the content of DNA, the approximation induced by the DBSCAN algorithm, and various factors that influence the formation of DNA damage, such as oxygen. The uncertainty associated with this parameter represents one of the primary sources of uncertainty in our simulation and model.

**Probability of cell death induced by the three cases**

Utilizing the formulas provided in Eq 5 and the parameter $p_{DNA}$=0.141 as specified in Table 1, we have fitted the survival fractions for different cell types to determine the probability of cell death caused by the three respective cases ($p_1$, $p_2$ and $p_3$). The goodness of the fitting is estimated using all datapoints of the survival fraction. The resulting probabilities are presented in Table 2. The cell survival fraction data for samples No. 1 through 16 were obtained by Suzuki et al. (*38*)., sample No. 17 was obtained from the study by Bronk et al. (*39*), samples No. 18 to 20 were sourced from the study by Furusawa et al. (*37*).

Table 2. Probability of cell death induced by the three cases through fitting and goodness of fitting

| No. | Cell type | Parameters (95% CI) | | | $R^2$ | Reference |
|---|---|---|---|---|---|---|
| | | $p_1$ | $p_2$ | $p_3$ | | |
| 1 | NB1RGB | 0.016 (0.012 - 0.020) | 0.061 (0.053 - 0.069) | 0.39 (0.15 - 0.62) | 0.9918 | (*38*) |
| 2 | HFL-III | 0.012 (0.010 - 0.014) | 0.060 (0.056 - 0.064) | 0.61 (0.50 - 0.73) | 0.9976 | (*38*) |
| 3 | LC-1 sq | 0.0087 (0.0044 - 0.013) | 0.033 (0.025 - 0.041) | 1.17 (0.94 - 1.41) | 0.9958 | (*38*) |
| 4 | A-549 | 0.0003 (-0.0026 - 0.0033) | 0.03 (0.030 - 0.036) | 0.46 (0.37 - 0.55) | 0.9957 | (*38*) |
| 5 | C32TG | 0.0082 (0.0053 - 0.011) | 0.046 (0.042 - 0.051) | 0.32 (0.18 - 0.46) | 0.9935 | (*38*) |
| 6 | Marcus | 0.0034 (-0.0050 - 0.012) | 0.047 (0.035 - 0.059) | 0.48 (0.15 - 0.81) | 0.9629 | (*38*) |
| 7 | U-251 MG (KO) | 2.37e-14 (-0.0085 - 0.0085) | 0.032 (0.017 - 0.048) | 0.70 (0.24 - 1.16) | 0.9762 | (*38*) |
| 8 | SK-MG-1 | 0.00094 (-0.0034 - 0.0053) | 0.033 (0.025 - 0.041) | 0.61 (0.36 - 0.85) | 0.9912 | (*38*) |
| 9 | KNS-89 | 2.24e-14 (-0.0044 - 0.0044) | 0.051 (0.044 - 0.058) | 0.84 (0.67 - 1.01) | 0.9927 | (*38*) |
| 10 | KS-1 | 0.025 (0.018 - 0.033) | 0.075 (0.060 - 0.091) | 0.27 (-0.22 - 0.77) | 0.9888 | (*38*) |
| 11 | A-172 | 0.011 (0.0016 - 0.020) | 0.068 (0.051 - 0.084) | 0.77 (0.27 - 1.28) | 0.9787 | (*38*) |
| 12 | ONS-76 | 0.0006 (-0.0093 - 0.011) | 0.062 (0.045 - 0.080) | 0.88 (0.34 - 1.42) | 0.9751 | (*38*) |
| 13 | KNS-60 | 2.25e-14 (-0.0034 - 0.0034) | 0.037 (0.032 - 0.042) | 0.44 (0.30 - 0.57) | 0.9936 | (*38*) |
| 14 | Becker | 0.0024 (0.00010 - 0.0046) | 0.020 (0.016 - 0.023) | 0.23 (0.16 - 0.30) | 0.9883 | (*38*) |
| 16 | T89G | 4.21e-12 (-0.0055 - 0.0055) | 0.025 (0.018 - 0.033) | 0.49 (0.28 - 0.71) | 0.9848 | (*38*) |
| 16 | SF126 | 0.0057 | 0.033 | 0.65 | 0.9951 | (*38*) |
| | | (0.0022 - 0.0092) | (0.027 - 0.040) | (0.46 - 0.85) | | |
| 17 | H460 | 1.42e-08 | 0.141 | 0.86 | 0.9756 | (*39*) |
| | | (-0.0081 - 0.0081) | (0.126 - 0.157) | (0.43 - 1.28) | | |
| 18 | V79 | 0.0040 | 0.060 | 0.19 | 0.9786 | (*37*) |
| | | (-0.0024 - 0.010) | (0.052 - 0.068) | (0.03 - 0.35) | | |
| 19 | HSG | 0.011 | 0.064 | 0.53 | 0.9996 | (*37*) |
| | | (0.0096 - 0.012) | (0.062 - 0.066) | (0.48 - 0.58) | | |
| 20 | T1 | 2.84e-13 | 0.074 | 0.48 | 0.9573 | (*37*) |
| | | (-0.011 - 0.011) | (0.059 - 0.090) | (0.16 - 0.80) | | |

The results show that $p_3 > p_2 > p_1$, indicating that radiation-induced cell death is related to the interactions between damaged DNA segments and the scale of potential structural variants. The result that case 2 and case 3 are more likely to induce cell death than case 1 may be attributed to the higher probability of misrepair associated with frequent interactions. In case 2, the DNA content in a TAD ranges from hundreds of kbp to several Mbp. These small-segment structural variants are less likely to induce cell death than the large-segment structural variants induced by case 3. For some cell types, the values of $p_3$ exceed 1.0, which can be explained by the difference in TAD - TAD interactions between the nuclear model we used and the actual cells. TAD-TAD interactions differ among cell types. For example, HeLa cells have 3,824 significant interactions, while IMR90 cells have 2,349 (*22*).

**Correlation between DSB distribution and cell death induced by high LET irradiation**

The simulation outcomes for C-12 ions with varying LET values are depicted in Figure 4. Since the incidence of DSBs is directly proportional to the dose, we have chosen to display the yield of DSBs ($k_{DSB}$). For the incidence of case 2 caused by a single particle and the incidence of case 1, we present the factors of the first-order terms ($a_{2,s}$ and $a_1$). Regarding the incidence of case 2 resulting from multiple particles and the incidence of case 3, due to the high statistical error associated with the parameters, we instead show the incidences at a dose of 1 and 2 Gy ($n_{2,m}(1 \text{ Gy})$, $n_{2,m}(2 \text{ Gy})$, $n_3(1 \text{ Gy})$ and $n_3(2 \text{ Gy})$).

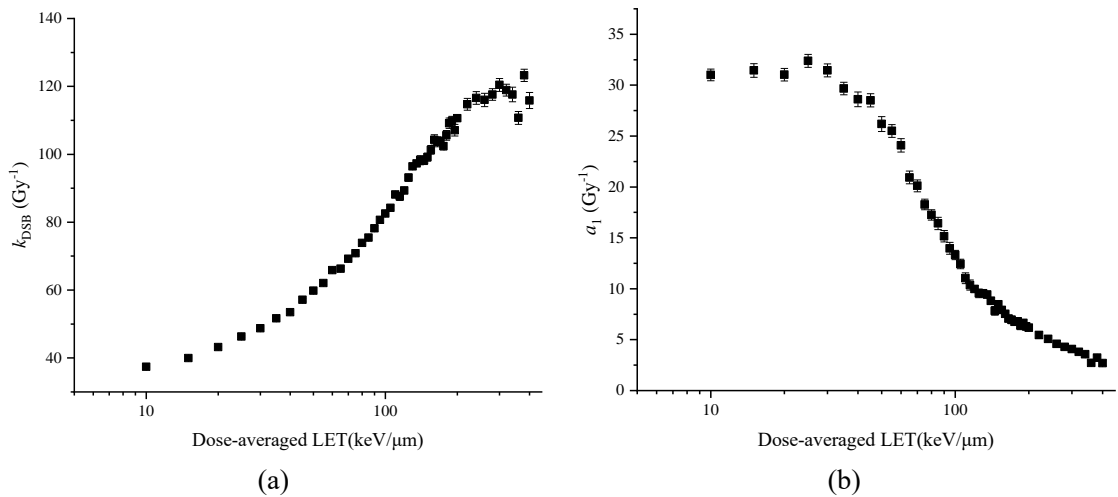

(a)  (b)

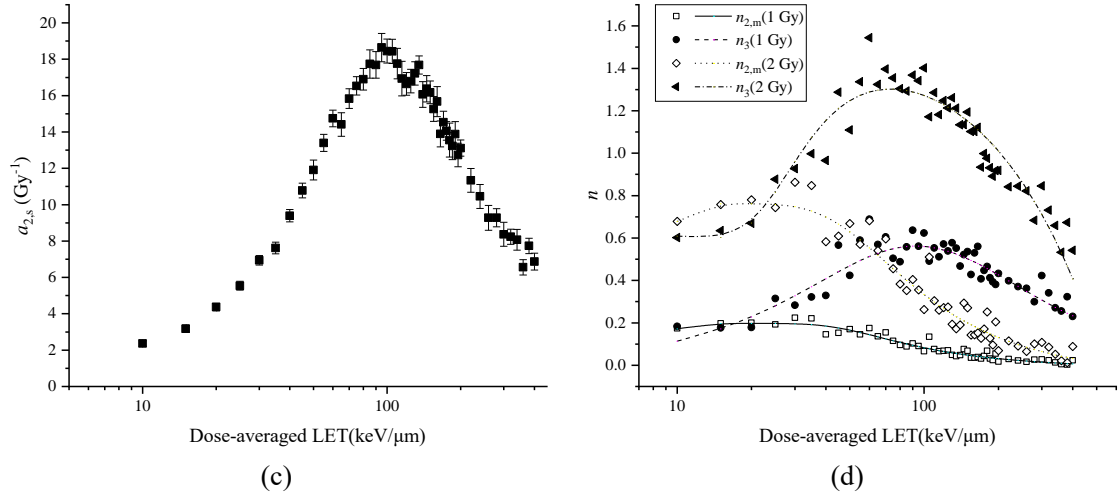

Figure 4. Simulation results of C-12 ions. (a) yield of DSB; (b) first-order factor of case 1; (c) first-order factor of case 2 induced by one particle; (d) incidence of case 2 induced by multiple particles and incidence of case 3 at 1 and 2 Gy.

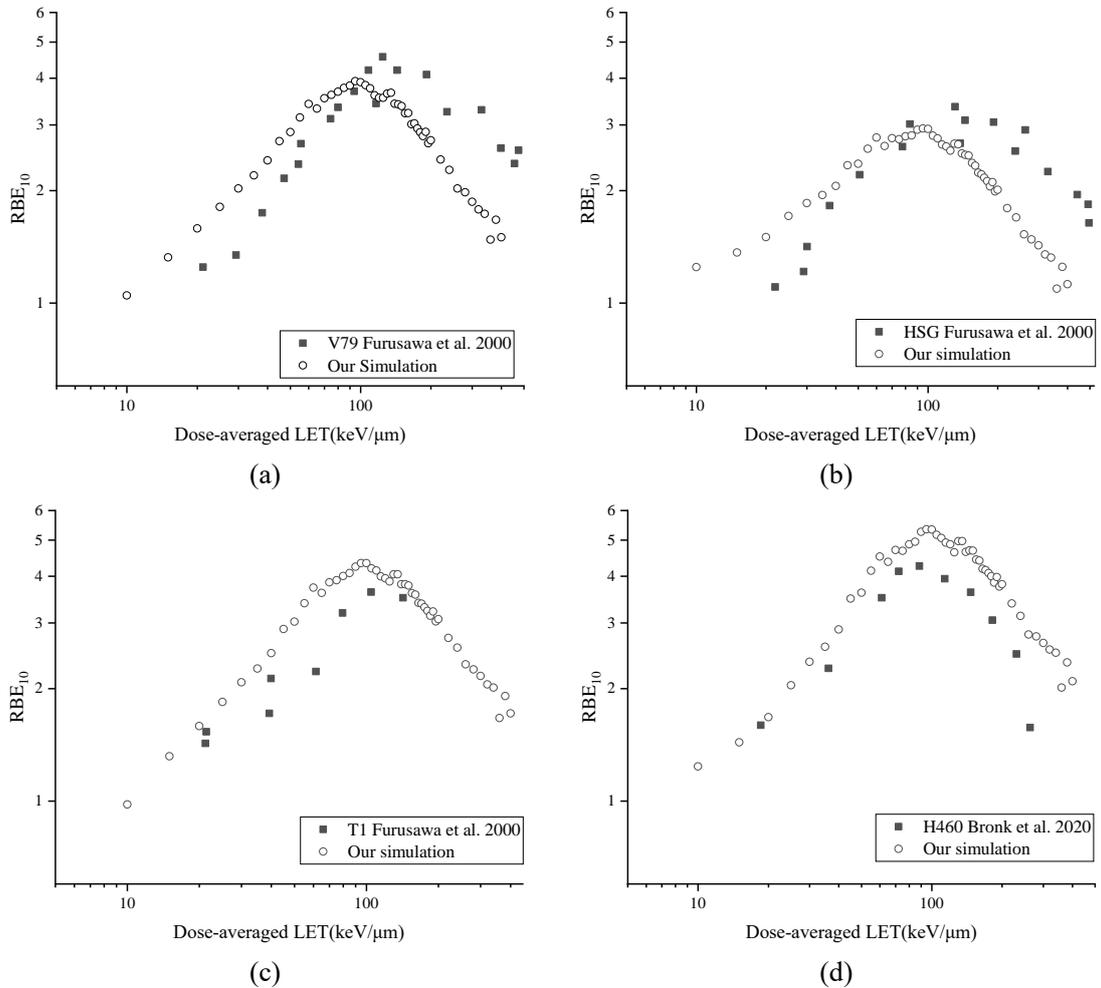

Figure 5. Comparison between the RBE at 10% of V79, HSG, T1 and H460 cells calculated using our simulation and the experimental results obtained by Furusawa et al. and Bronk et al.

The results indicate that the yield of DSBs increases with the increase in LET. This trend corresponds to the simulation results of other codes (*12*). In comparison to the single-particle-induced case 2 for high LET irradiation, the contribution of case 2 induced by multiple particles is negligible. The yield of case 2 reaches its maximum at an LET of around 100 keV/μm. The incidence of case 3 shows a peak within the LET range from tens of keV/μm to about 100 keV/μm, as can be observed from the trending lines.

Moreover, by utilizing the probabilities of cell death resulting from the three cases (as obtained from low LET irradiation), we have calculated the relative biological effectiveness (RBE) for C-12 ion irradiation. The calculated RBE at 10% survival fraction of V79, HSG, T1 and H460 cells are presented and compared with experimental results provided by Furusawa et al. (*37*) and Bronk et al. (*39*) in Figure 5.

The trend of our simulation results for RBE versus LET roughly aligns with the experimental results. Empirical RBE data suggests that RBE approximately peaks when the LET is around 100 - 200 keV/μm (*40*). This consistency further indicates a correlation between the distribution of DSBs in the 3D genome and the subsequent induction of cell death by high LET irradiation. It should be noted that there is still a systematic discrepancy between our simulation results and the experimental results. This may be attributed to the approximations introduced by track-structure simulation, the DBSCAN algorithm for identifying DSB, and the differences in TADs and nuclear models. These factors represent the limitations of our study.

**Correlation between DSB distribution and radiation quality**

Radiological risk is closely associated with radiation-induced DNA damage. Our hypothesis proposes that DSBs occurring in DNA segments that frequently interact are more likely to result in misrepair. We further speculate that the distribution of DSBs within the 3D genome structure is also related to radiation-induced stochastic effects. In the low-dose range that is of concern in radiation protection, we focus on the events induced by a single particle. In the mathematical formulation, the first-order deviation at 0 Gy (which approximates the first-order term factor) is related to these events. The radiation quality factor ($Q$) is a metric for quantifying the stochastic effects elicited by high LET irradiation in contrast to low LET irradiation. Its values are derived by collating RBE results associated with radiation-induced mutations and cancers. Different organizations, such as the ICRP and NASA, adopt different considerations, data-processing methods, and data sources, thereby arriving at different values. In Figure 6, we have presented the ratio of the first-order factor of high LET irradiation to that of low LET irradiation ($a_{2,s}$(carbon)/$a_{2,s}$(electron)), as well as the first-order factor of case 3 ($a_3$), along with the $Q$ recommended by ICRP 92 (*41*) and as derived by Borak et al. (*42*) (based on NASA's model), to illustrate their correlations.

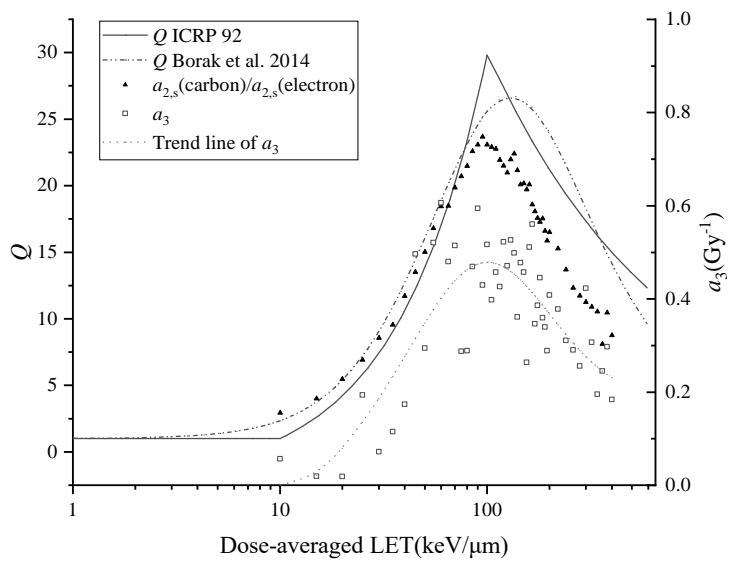

Figure 6. Correlation between the Distribution of DSBs in the 3D genome and radiation quality. The ratio of the first-order factor of carbon ion irradiation to that of electron irradiation ($a_{2,s}$(carbon)/$a_{2,s}$(electron)) is depicted on the left axis and the first-order factor of case 3 ($a_3$) is depicted on the right axis.

Both the curves of case 2 and case 3 varying with LET are similar in shape to the curve of the $Q$ value changing with LET (the curves present a peak at around 100 keV/μm), indicating that case 2 and case 3 are correlated with the $Q$ value. The consequence of case 2 is small-segment structural variants. This type of mutation is less likely to induce cell death than the large-segment structural variants induced by case 3. Therefore, cells carrying this mutation may be more likely to survive and give rise to cancer and genetic effects.

**Discussion on the "target" and "domain"**

The identification of physical and biological entities for "targets" or "domains," is a crucial process for uncovering the mechanisms underlying radiobiological cell death. The "targets" or "domains" should be able to explain at least the "shoulder" phenomenon observed in the survival fraction curves of low LET irradiation and the relationship between RBE and LET. A wealth of phenomena indicate that DSBs alone do not fully represent the concept of a "target" or "domain". Theories that rely solely on this assumption are inadequate for accurately predicting cell survival fractions. The hypothesis based on complex DSBs at the scale of tens of bp is also difficult to explain the "shoulder" phenomenon. This is because the probability of multiple particles hitting a region of several nanometers at a dose of several Gy is too low to account for the "shoulder" (*10*). Moreover, researchers have found that the clustering of DNA breaks can serve as a predictor of cell death under high LET irradiation and have discussed the potential role of TADs in the radiobiological effect (*43*, *44*). These findings indicate that the "domains" are on a larger scale than the nanoscale of DSBs. Recent theories have proposed the existence of a "domain" at a larger scale, with fitting results suggesting a diameter of approximately 500 nm for these domains (*15*, *16*). However, this scale does not correspond to any known cellular structure. TADs have a sub-micrometer size. Moreover, the sizes of TADs vary greatly. If TADs were the "domains" or "targets", the assumption of a single-sized domain would not be appropriate. Our simulation results show that DSBs in TADs with frequent interactions are more likely to induce cell death than clustered DSBs within a single TAD. The variability in the sizes of TADs and our study suggests that the concept of "targets" or "domains" as distinct physical or biological entities may be an oversimplification.

Besides the domain, e.g. TAD, the critical factor could be the interaction between damaged DNA segments and their distance in the genome. TADs and other architectural elements within the 3D genome are indeed characterized by the interactions among DNA segments. Therefore, a comprehensive model of radiobiological response should encompass the impact of these interactions and the spatial arrangement of DNA on radiobiological outcomes, rather than focusing solely on specific, fixed-size targets. However, due to the potentially complex relationship between the frequency of DNA segment interactions and cell death, our study employed a simplified model. In this model, the interactions between DNA segments are elegantly captured by representing them with TADs and those TADs that engage in frequent interactions.

**Discussion on the limitations of our study**

The objective of our study is to elucidate the correlation between the distribution of DNA within the 3D genome structure and radiation-induced cell death, rather than to develop a model for predicting cell survival at specific doses and LET values. There are several points that should be improved for future study.

Whether TADs reflect the statistical frequencies of chromatin interactions within a cell population or represent genuine physical units in each cell nucleus remains a matter of debate. Experiments have observed that repressed TADs form sphere-like structures (*45*). Single-cell studies suggest that such structures do exist, but there is significant variability within the cell population. Different cells have distinct geometries and 3D genome structures, which consequently may potentially influence the number and distribution of DSBs in the 3D genome. Our simulation only uses one nuclear model. However, the differences in sizes of TADs and relative interactions among cell types are not so great, so the correlation indicated by our study is still valid. Moreover, the intrinsic uncertainties associated with track-structure Monte Carlo simulations introduce significant uncertainty in the estimated quantities of DNA damage. The imprecise cross-section of high-energy ions and the approximations introduced by the DBSCAN algorithm when identifying DSBs contribute to this uncertainty. As demonstrated in the previous sections, the parameters in the simulation exert a profound impact on the results. However, the exact values of these parameters still remain uncertain. Our model simplifies DNA segment interactions into TADs and assumes the same probability within the same case while neglecting the variability among TADs. This limitation restricts its utilization in accurately predicting the cell survival fraction.

The motivation behind our study is to demonstrate the correlation between the distribution of DSBs in the 3D genome and the radiobiological effect, rather than to model the RBE for high LET irradiations. The simplified model is used to quantitatively indicate this correlation. We did not attempt to adjust the size of TADs, the parameters of the DBSCAN algorithm, and other adjustable parameters to fit all the experimental data. Our simulation can illustrate general trends rather than provide exact numerical values. The impact of DSB distribution in the 3D genome on the radiobiological effect has not been directly demonstrated through experiments. Experimental data on cell survival fractions often involve large uncertainties, which in turn restrict the derivation of a definite quantitative relationship for accurately predicting cell survival fractions. Over-tuning the parameters to fit the experimental results may mislead other researchers.

## Conclusion

This study proposes a hypothesis that the distribution of DSBs within the 3D genome is a predominant factor influencing radiation-induced cell death. We suggest that clustered DSBs in DNA segments with high interaction frequencies are more likely to result in cell death than isolated DSBs. TAD can be regarded as the reference unit for evaluating the impact of DSB clustering in the 3D genome. To study this correlation quantitatively, we developed a simplified model. In this model, DSBs are categorized into three cases based on their distributions across TADs. The incidences of each case of DSB induced by irradiation at various doses and LETs were calculated using track-structure Monte Carlo simulations. The simulation focuses on electrons and carbon ions to represent low and high LET irradiations. The results of fitting data from several cell types indicate that the probabilities of inducing cell death in the three cases follow this relationship: $p_3 > p_2 > p_1$ (DSBs in TADs with frequent interactions are significantly more likely to induce cell death than clustered DSBs within a single TAD. Case 2 is significantly more likely to cause cell death than isolated DSB). The curves depicting the incidence in case 2 and case 3 as a function of LET values show a shape similar to that of the radiation quality factor ($Q$) used in radiation protection. This suggests that these two cases are also related to the stochastic effects induced by high LET irradiation. Our study underscores the importance of the 3D genome structure in the mechanisms of radiobiological effects. Furthermore, it provides insight for subsequent studies investigating the role of the 3D genome and for the development of models that predict cell survival and RBE based on underlying mechanisms.

## Acknowledgments

This work was supported by the National Key Research and Development Program of China (Grant No. 2024YFA1014103), the National Natural Science Foundation of China (Grant No. 12405359) and the Postdoctoral Fellowship Program of Chinese Postdoctoral Science Foundation under Grant Number GZB20240342. The authors would like to express their gratitude to Dr. Wei Bo Li of the Federal Office for Radiation Protection (BfS), Germany, for his valuable suggestions.

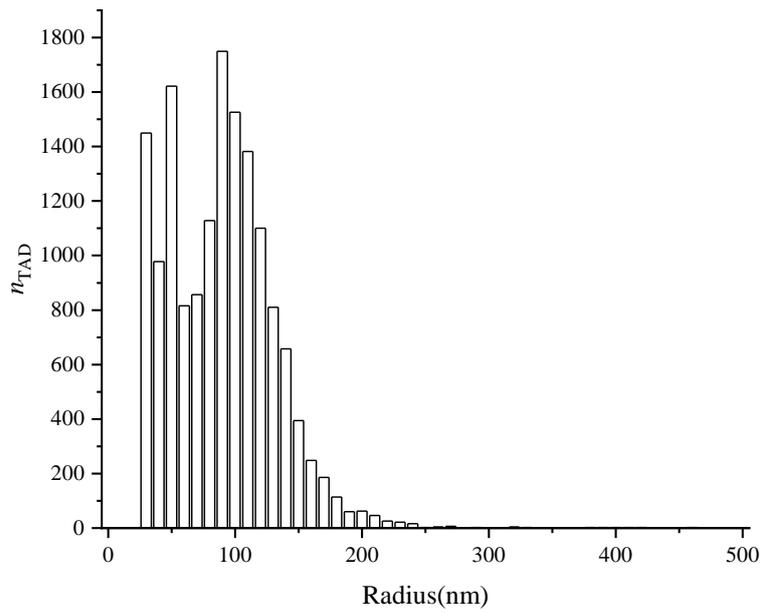

Figure S1. Histogram of radius of TADs

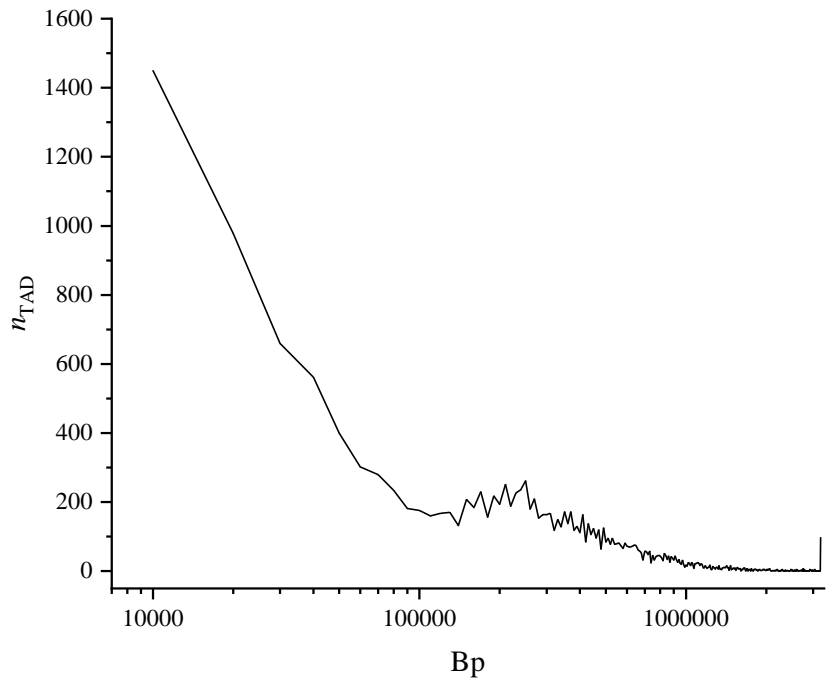

Figure S2. Histogram of DNA content of TADs

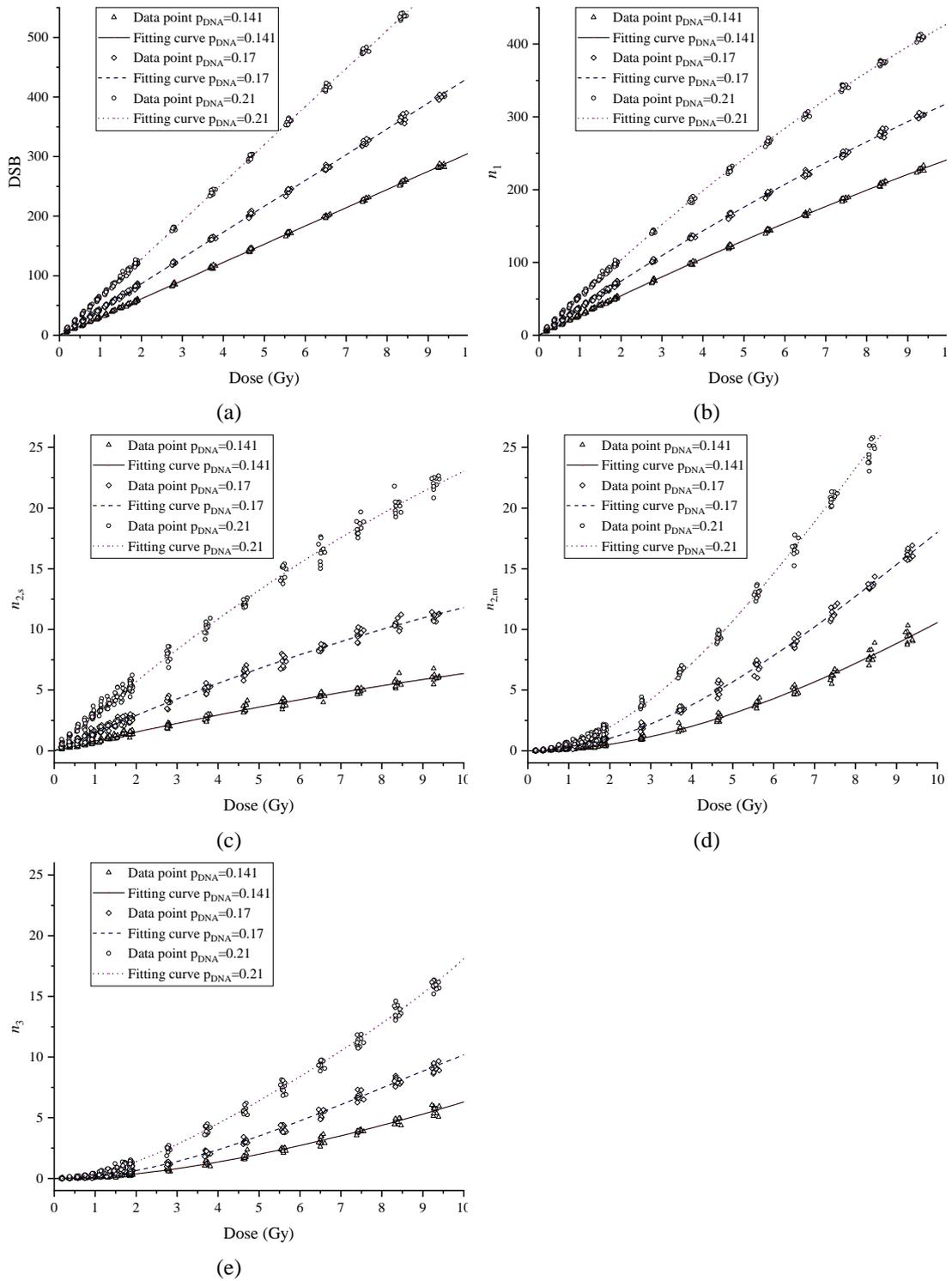

Figure S3. Incidence of events induced by electron irradiation with different $p_{DNA}$ values. (a) DSB; (b) isolated DSB (case 1); (c) clustered DSB in a single TAD (case 2) induced by one particle; (d) clustered DSB in a single TAD (case 2) induced by multi particles; (e) DSBs in TADs with frequent interactions (case 3).

Table S1. Parameters of formulas describing the incidence of three cases (with 95% CI)

| Parameters | $p_{DNA}$=0.141 | $p_{DNA}$=0.17 | $p_{DNA}$=0.21 |
|---|---|---|---|
| $k_{DSB}$ | 30.59<br>(30.54 - 30.64) | 43.30<br>(43.22 - 43.38) | 63.98<br>(63.88 - 64.08) |
| $a_1$ | 27.8<br>(27.70 - 28.00) | 38.4<br>(38.27 - 38.71) | 54.09<br>(53.83 - 54.35) |
| $b_1$ | 0.3665<br>(0.3462 - 0.3868) | 0.655<br>(0.626 - 0.6858) | 1.11<br>(1.082 - 1.154) |
| $a_{2,s}$ | 0.8009<br>(0.7785 - 0.8233) | 1.527<br>(1.495 - 1.559) | 2.988<br>(2.935 - 3.041) |
| $b_{2,s}$ | 0.01646<br>(0.01343 - 0.01949) | 0.03458<br>(0.03031 - 0.03885) | 0.06842<br>(0.06139 - 0.07545) |
| $b_{2,m}$ | 0.1396<br>(0.1338 - 0.1454) | 0.2744<br>(0.2680 - 0.2808) | 0.5342<br>(0.5240 - 0.5444) |
| $c_{2,m}$ | 0.003387<br>(0.002682 - 0.004092) | 0.009434<br>(0.008658 - 0.01021) | 0.02132<br>(0.02008 - 0.02256) |
| $b_3$ | 0.1081<br>(0.0954 - 0.1208) | 0.1818<br>(0.1661 - 0.1975) | 0.4258<br>(0.4036 - 0.4480) |
| $c_3$ | 0.006831<br>(0.003222 - 0.01044) | 0.008832<br>(0.004414 - 0.01325) | 0.0434<br>(0.03722 - 0.04974) |
| $d_3$ | 0.0002325<br>(-0.0000163 - 0.0004813) | 8.33E-05<br>(-0.0002134 - 0.00038) | 0.00190<br>(0.001469 - 0.002333) |